  \documentclass[manuscript]{aastex}

\usepackage{enumerate}

\shorttitle{RHESSI and SDO/AIA plasma parameters}
\shortauthors{M.B. \& E.P.K.}
\begin{document}

\title{RHESSI and SDO/AIA observations of the chromospheric and coronal plasma parameters during a solar flare}
\author{M. Battaglia \altaffilmark{1,2}, E. P. Kontar \altaffilmark{2}}
\affil{Institute of 4D Technologies, School of Engineering, University of Applied Sciences and Arts Northwestern Switzerland, 5210 Windisch, Switzerland}
\affil{SUPA, School of Physics and Astronomy, University of Glasgow, G12 8QQ, UK}
\email{e-mail: marina.battaglia@fhnw.ch}

\begin{abstract}
X-ray and EUV observations are an important diagnostic of various plasma parameters of the solar atmosphere during solar flares. Soft X-ray and EUV observations often show coronal sources near the top of flaring loops, while hard X-ray emission is mostly observed from chromospheric footpoints. Combining RHESSI with simultaneous SDO/AIA observations, it is possible for the first time to determine the density, temperature, and emission profile of the solar atmosphere over a wide range of heights during a flare, using two independent methods. Here we analyze a near limb event during the first of three hard X-ray peaks. The emission measure, temperature, and density of the coronal source is found using soft X-ray RHESSI images while the chromospheric density is determined using RHESSI visibility analysis of the hard X-ray footpoints. A regularized inversion technique is applied to AIA images of the flare to find the differential emission measure (DEM). Using DEM maps we determine the emission and temperature structure of the loop, as well as the density, and compare it with RHESSI results.
The soft X-ray and hard X-ray sources are spatially coincident with the top and bottom of the EUV loop, but the bulk of the EUV emission originates from a region without co-spatial RHESSI emission.
The temperature analysis along the loop indicates that the hottest plasma is found near the coronal loop top source. The EUV observations suggest that the density in the loop legs increases with increasing height while the temperature remains constant within uncertainties. 
\end{abstract}
\keywords{Sun: flares -- Sun: X-rays, $\gamma$-rays -- Sun: Chromosphere}
\maketitle
\section{Introduction}
Solar flares are one of the most spectacular solar phenomena and are observed over a wide range
of the electromagnetic spectrum from radio frequencies to high energy gamma-ray emission.
These solar explosions are associated with the acceleration of large numbers of energetic electrons,
plasma heating, and plasma motions. However, the details of plasma heating and
particle acceleration and transport in solar flares are still poorly understood.
Over the last decade, spatially resolved hard X-ray (HXR) observations with RHESSI \citep{Li02} have notably
enriched our understanding of solar flares. Using these X-ray observations the spatial,
angular, and energy characteristics of non-thermal electrons in solar flares have been
deduced \citep[see][as a recent review]{2011SSRv..159..301K}, which resulted in improved
understanding of non-thermal electron evolution \citep[see][as a recent review]{2011SSRv..159..107H}.
The detailed analysis of the footpoint emission, the brightest HXR sources in solar flares,
revealed the density structure of the chromosphere \citep{Ba11a,Ba11b}.
At the same time RHESSI provides diagnostics of thermal plasma in the soft X-ray (SXR) emitting loop.
This allows for measuring the temperature, emission measure, and density in the corona,
namely the often observed loop-top coronal SXR sources \citep[e.g.][]{Ba09,Ve05}.

However, RHESSI observations are limited to electrons with plasma temperatures higher than around ~8~MK. Since the plasma in a flare is unlikely to be isothermal and the temperature is expected to vary over time and for different flare sources one has to extend the observed temperature range to lower values such as observed in EUV. For example, Hinode spectroscopic studies of hot plasmas provided details of plasma motions
\citep[e.g.][]{2008ApJ...685.1277W,2009ApJ...696..121L,2010ApJ...719..213W}
and observations of decaying post-flare loops \citep[e.g.][]{1995ApJ...446..860F,2000A&A...355..769V}
provide estimates of the electron density of the bright regions at the loop tops.
Line spectroscopy can provide information on the emission measure at different temperatures \citep[eg.][]{2011A&A...532A..27G,2009A&A...494.1127R}, while \citet{2011ApJ...734...34K} showed how diffraction patterns in TRACE 171 \AA\ images can be used to make temperature and density estimates.
Since its launch in 2010, SDO/AIA \citep{Le11} has been contributing greatly to our knowledge of the solar flare EUV emission
via full disk imaging allowing ready comparison with X-ray observations of flaring events.
It covers a temperature range from $\sim$5000~K to $\sim$20~MK in 9 wavelengths and,
even more importantly, provides the spatial resolution allowing to connect the non-thermal
and thermal phenomena in solar flares. Combining RHESSI and SDO/AIA observations of flaring loops
(complemented by GOES SXR observations) thus allows to analyze the temperature and density structure from the chromosphere to the corona
during a solar flare using two complementary methods. 

In this paper we present simultaneous RHESSI, SDO/AIA, and GOES observations taken during the first of three
HXR peaks of a limb flare. Fitting RHESSI full Sun spectra with a single temperature model provides the SXR temperature and emission measure while RHESSI images indicate the location of the hot plasma.
This is enriched by temperature and emission measure found from GOES, which is most sensitive
to temperatures around the RHESSI sensitivity range and down to a few MK. HXR footpoint visibility analysis
gives information about the chromospheric density structure while SDO/AIA observations
are used for differential emission measure (DEM) analysis along the flaring loop. For the DEM analysis a regularized inversion method, developed by \citet{2012A&A...539A.146H}, was used. This allows to make full use of the information contained in the 6 AIA wavelengths sensitive to coronal temperatures as opposed to using filter ratios or single temperature fits. 
\section{Data analysis}
We consider a well observed limb-flare that happened on 2011 February 24 with three HXR peaks between 07:29
and 07:33 UT. This GOES M3.5 class flare has one well defined footpoint during the first HXR peak, the height structure
of which was analyzed in some detail by \citet{Ba11b}. The flare has demonstrated white-light emission
and is also associated with a large filament eruption.\footnote{http://www.astro.gla.ac.uk/users/mbattaglia/20110224\_online\_material/}

\begin{figure*}
\begin{center}
\includegraphics[width=7cm]{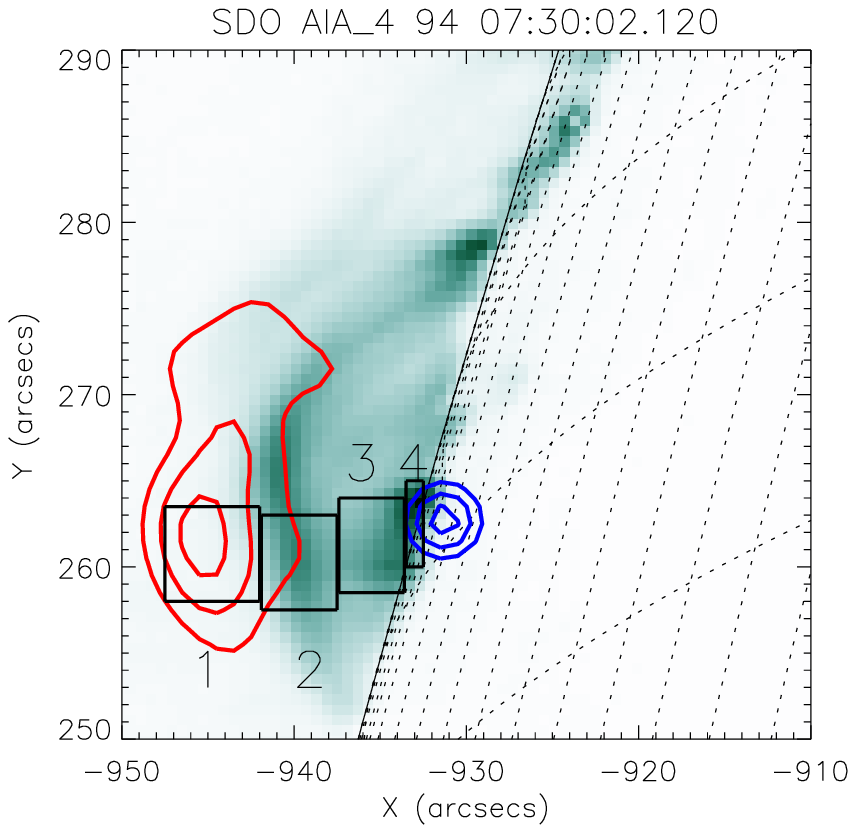}
\includegraphics[width=7cm]{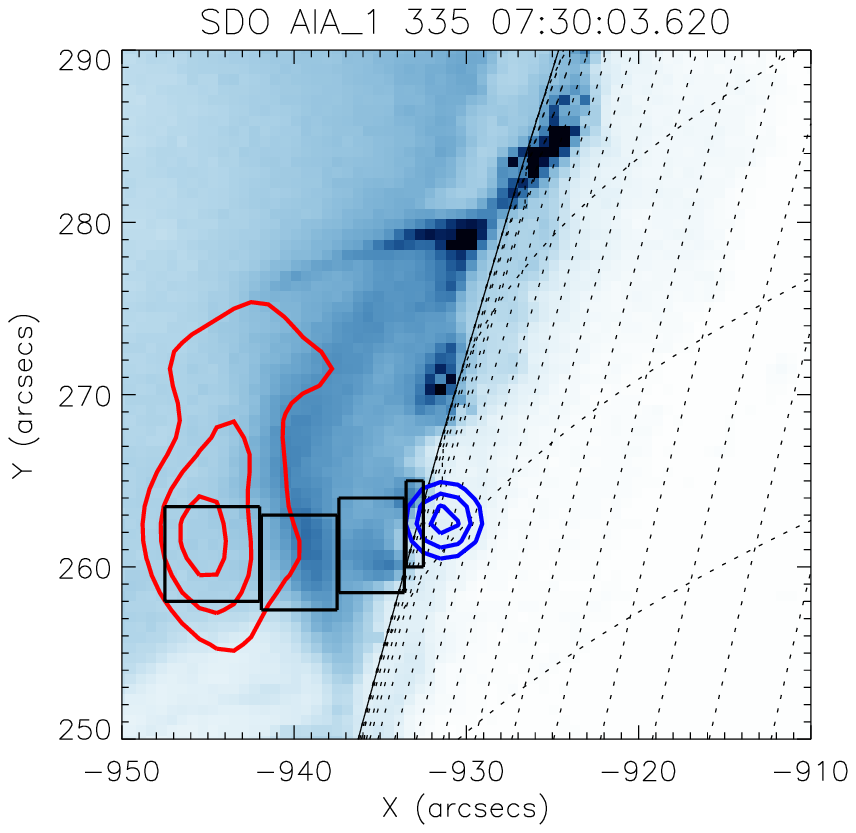}\\
\includegraphics[width=7cm]{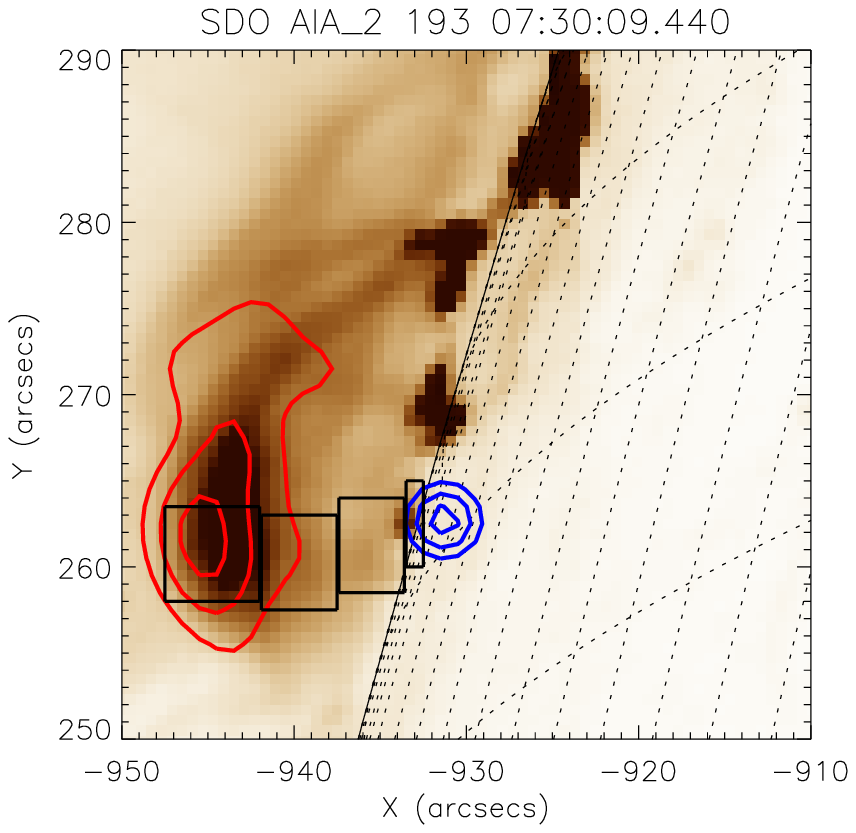}
\includegraphics[width=7cm]{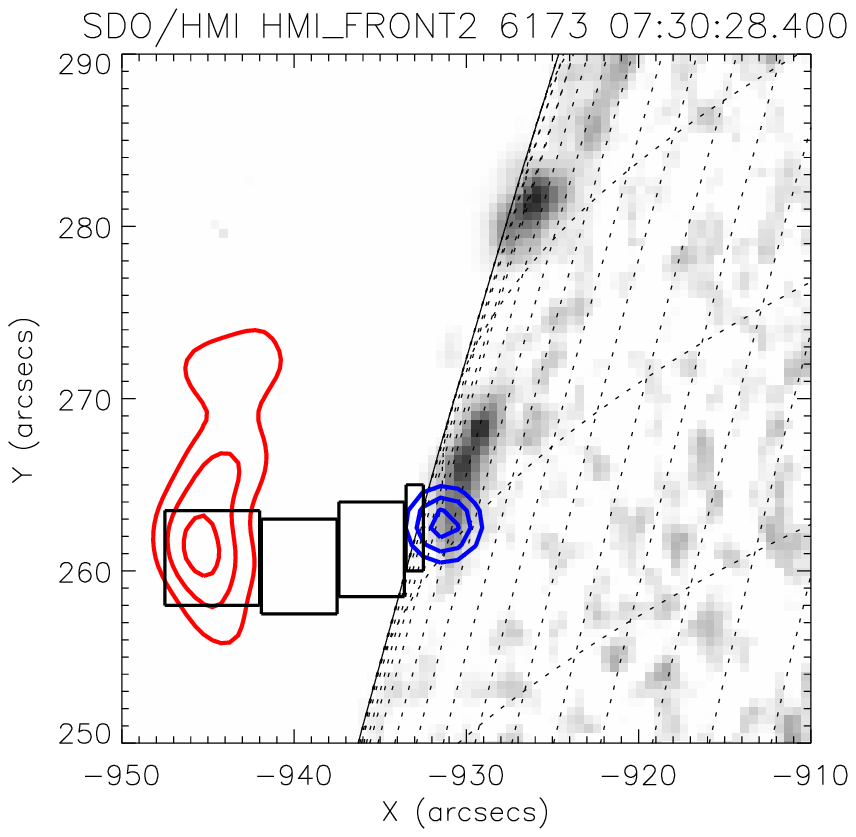}
\end{center}
\caption {Top left: AIA 94 \AA\ image overlaid with RHESSI thermal emission (red; 50, 70, 90 \% contours in CLEAN image)
and non-thermal emission (blue; 50, 70, 90 \% contours of circular Gaussian visibility fit).
The rectangles indicate the regions of interest on which the analysis is focused.
Top right: corresponding 335 \AA\ image.
Bottom left: Respective AIA 193 \AA\ image showing saturation near the coronal source. Bottom right: HMI difference white light image. }
\label{aia94}
\end{figure*}

In the present study we focus on the time of the first HXR peak, around 07:30 UT, because of
substantial saturation of AIA images at later times.  The spatial structure of the flare as
observed by both RHESSI above $\sim6$~keV and AIA in the 94~\AA, 193~\AA, 335~\AA ,
and Helioseismic and Magnetic Imager \citep[HMI,][]{2012SoPh..275..207S} white light 6173~\AA\ continuum channels is shown in Figure \ref{aia94}.
The analysis is focused on four physically distinct regions of interest,
namely the coronal source, two parts of the flaring magnetic loop,
and the HXR footpoint. Rectangles in Figure \ref{aia94}
show the corresponding regions selected for the analysis.
\subsection{Alignment between AIA images and RHESSI}
The co-alignment between AIA images in different wavelength channels is important for the accurate
determination of the DEM while correct overlays of AIA images and RHESSI images are necessary
to determine co-spatial X-ray and EUV sources. 
The alignment between the different wavelength channels of AIA was determined using the
limb-fitting routine \textit{aia\_coalign\_test} by \citet{2011SoPh..tmp..384A}. This gives
an average co-alignment accuracy of $1.1\pm0.6$ pixels.
The co-alignment between RHESSI and AIA images has been checked
using the method described in \citep{Ba11b}. Solar Aspect System and Roll Angle System provide
pointing knowledge \citep{2002SoPh..210...87F} so that the RHESSI disk center is known
to a degree of accuracy better than $0.2$~arcsec.
The disk center of SDO images can also be accurately verified using AIA full disk images in the chromospheric wavelengths (1600~\AA\ \& 1700~\AA) and WL images from HMI. This cannot be done for the other wavelength channels since the very non-uniform distribution of EUV emission in active regions introduces a strong bias. 
The SDO position of the solar-disk center is found to deviate by only 0.4/0.6~arcsec
in the $x/y$ direction from the RHESSI disk center for the time interval analyzed here.
The larger pointing uncertainty is related to the absolute calibration of roll-angle of the SDO images
(rotation around the disk center). Comparison of similar features and
the position of the HXR footpoints relative to the 1700~\AA\ and 1600~\AA\ emission suggests
an uncertainty of around 0.2 degrees which corresponds to $\sim 0.2\degr 960''\pi/180\degr=3.2''$ uncertainty
in the tangential direction near the limb. However, the radial direction uncertainty is less than $0.6''$ allowing us to have rather precise measurements for near limb events. 
Since we are investigating regions of interest of several arcsecond diameter,
these uncertainties in relative pointing should not affect the analysis, thus RHESSI images
were not shifted nor rotated relative to AIA since any such manipulation
would be subjective.\footnote{The SDO images presented in Figure \ref{aia94} can be rotated
to make RHESSI, EUV, and WL images coincide, which is subjective and affects its tangential alignment
but not the radial.}
\subsection{AIA emission measures and temperatures using regularized inversion}
Full Sun AIA level 1 images were prepped to level 1.5 and normalized by exposure time \citep[see][and online documentation linked therein]{Le11}.
Using only AIA wavelength channels with un-saturated images sensitive to coronal temperatures we find
the line-of-sight DEM from AIA data using the method of regularized
inversion \citep{2012A&A...539A.146H}. The method allows to reconstruct the DEM, $\xi (T)$, for a selected
area or individual pixel,
\begin{equation}\label{eq:dem_def}
   \xi (T)=n^2\frac{dl}{dT} \;\;\;\; [\mbox{cm}^{-5}\,\mbox{K}^{-1}]
\end{equation}
where $n$ is the plasma density, $T$ is the temperature and $l$ is the distance along the line of sight.

\begin{figure*}
\begin{center}
\includegraphics[height=7cm]{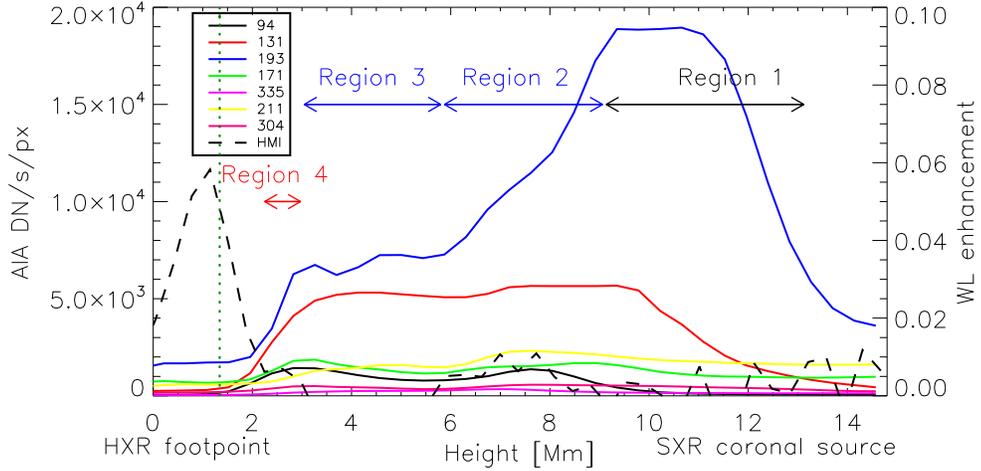}
\end{center}
\caption {Profiles of AIA maps in different wavelengths (see legend) as a function of absolute height above the photosphere as found from HXR visibility analysis (see text). The arrows indicate the extent in x-direction of the regions of interest that were analyzed. The green dashed line indicates the height of the HXR peak emission at 30-40 keV. The black dashed line gives the HMI WL enhancement.}
\label{saturation}
\end{figure*}

Using a set of different wavelength values from a pixel or a region in AIA maps,
the method returns a regularized DEM as a function of $T$. This allows unambiguous determination of the contribution
of different temperatures to the total emission measure as opposed to RHESSI where the spectrum is often consistent with different combinations of EM/T \citep[e.g.][]{2006SoPh..237...61P}. Due to high photon flux during flares
images in some of the wavelengths tend to be saturated during the main phase of the flare so that not all AIA wavelength channels can be used to find the DEM during the peak of the flare.
In the present flare the image in the 131 \AA\ wavelength channel was saturated for most
of the flaring loop region, while the 193 \AA\ image saturated near the top of the coronal source.
This is illustrated in Figure \ref{saturation} showing the AIA image profiles along the x-direction averaged over 6 arcsec in the y-direction. This one-dimensional cut across the images (Figure~\ref{aia94})
includes the flaring footpoints, the legs of the magnetic loop, and the coronal source.

Therefore, for the subsequent DEM analysis, we use images at 94, 171, 335, 193,
and 211~\AA\ for the EUV-loop and HXR footpoint regions, omitting 193~\AA\ for the region
at the loop-top where the image in this wavelength was saturated. We discuss the consequences of this omission in Section \ref{secsaturation}.
For the inversion we use $0$-th order regularization in a range of temperatures  $\log_{10}T=5.7$
to $\log_{10}T=7.5$ with 16 temperature bins. This corresponds to $\mathrm{\Delta logT}=0.11$  in temperature space.
The latest version of the AIA response function (including the new CHIANTI fix \footnote{January 2012, http://sohowww.nascom.nasa.gov/solarsoft/sdo/aia/response/chiantifix\_notes.txt}) was used.
\section{DEM analysis of selected spatial regions} \label{secdem}
\begin{figure*}
\begin{center}
\resizebox{0.95\hsize}{!}{\includegraphics{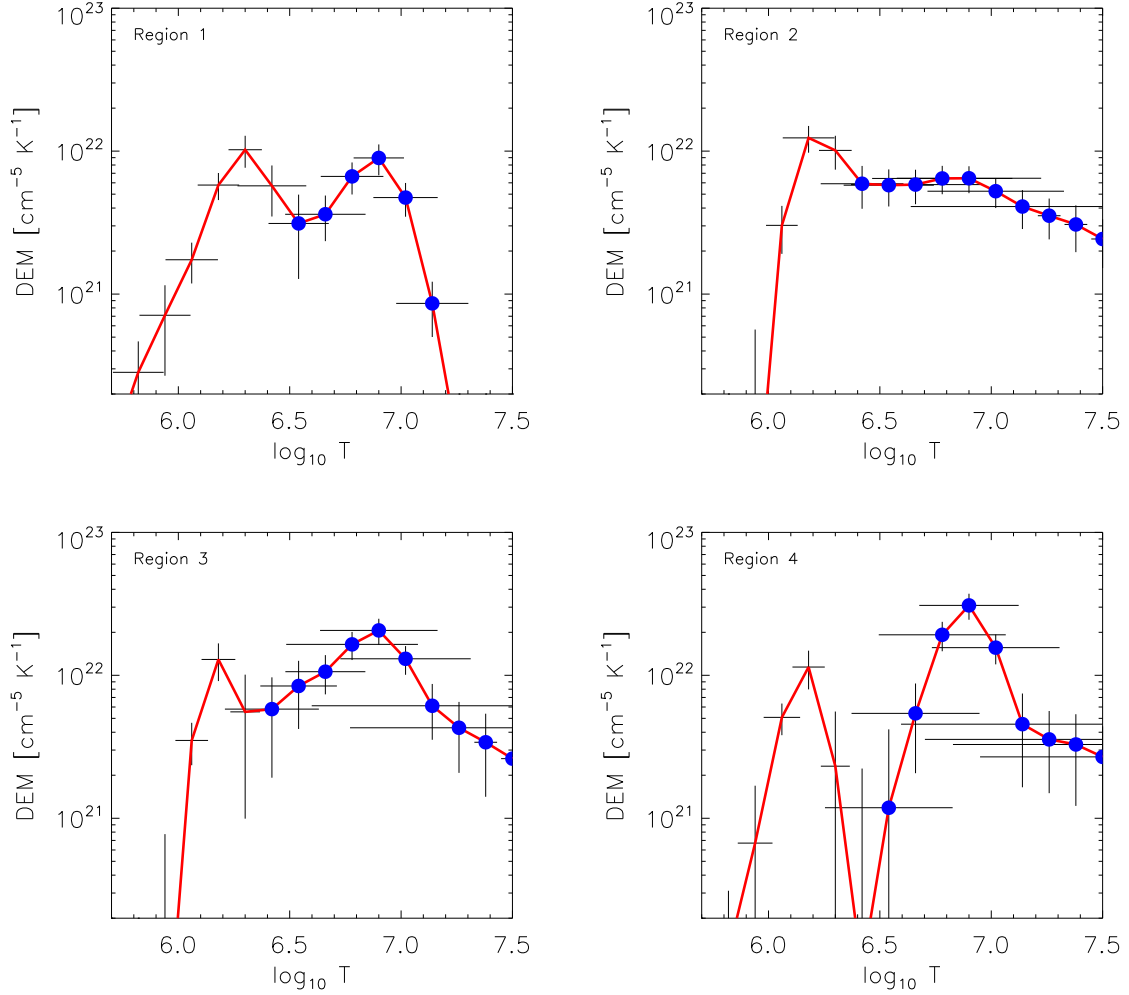}}
\end{center}
\caption {DEM within the regions of interest indicated in Figure \ref{aia94} as a function of $T$, using the 94, 171, 211, and 335 \AA\ wavelength channels (top left) and 94, 171, 211, 335, and 193 \AA\ (top right to bottom right, see Section \ref{secsaturation}). The blue dots give the temperatures over which the DEM was integrated to find the total emission measure (see Section \ref{secdem}).}
\label{totalarea}
\end{figure*}

\begin{figure*}
\begin{center}
\includegraphics[width=14cm]{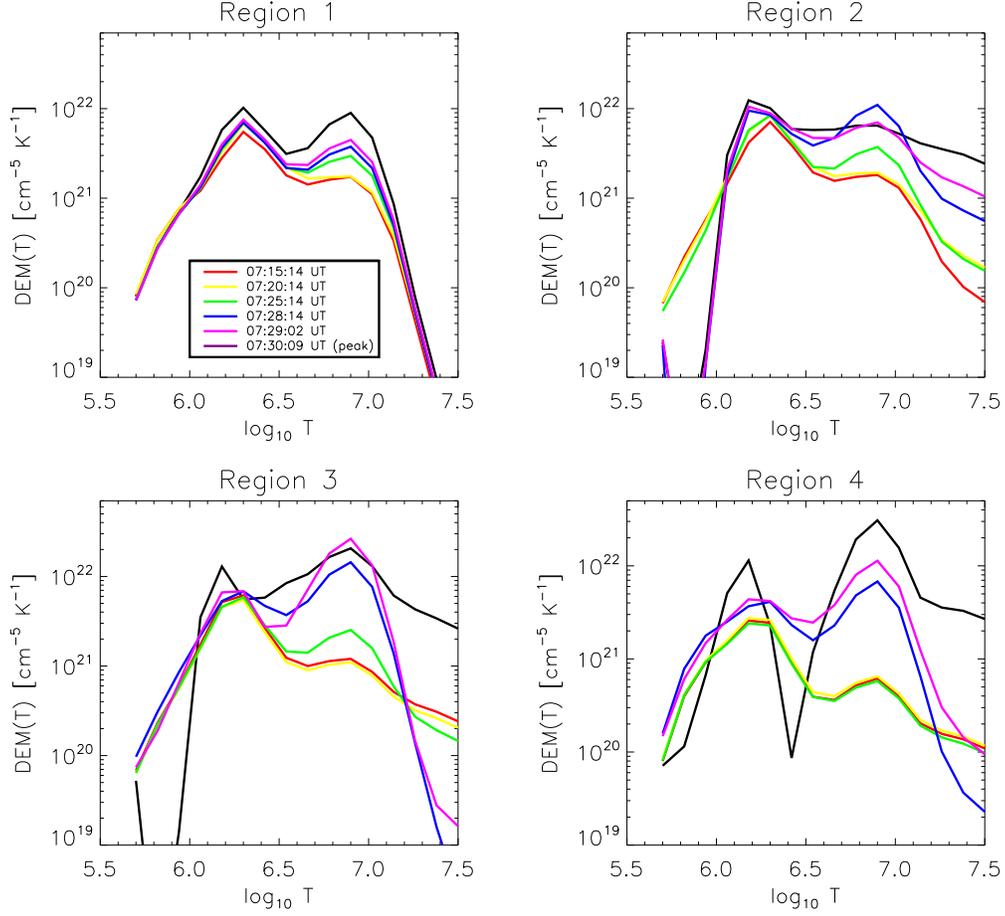} 
\end{center}
\caption {Pre-flare time evolution of DEMs, using the 94, 171, 211, and 335 \AA\ wavelength channels (top left) and 94, 171, 211, 335, and 193 \AA\ (top right to bottom right, see Section \ref{secsaturation}). The black line is the DEM during the main analysis interval, the colored lines give the DEM at different 
times (from about 15 min to 1 min) before the flare.}
\label{preflaredem}
\end{figure*}

We focus the analysis of emission measure, temperature, and density on a few selected,
physically meaningful regions:
\begin{enumerate}[i)]
\item Coronal source, corresponding to the bright SXR source at the top of the loop (region 1 in Figure \ref{aia94})
\item EUV loop. Two regions are selected at the southern loop leg between the SXR coronal source and the footpoints (regions 2 and 3)
\item  Footpoint region. This region is near the origin of the HXR and white light emission but also contains some hot plasma observed in EUV (region 4 in Figure \ref{aia94}).
\end{enumerate}
For each of the selected regions we find the total DEM from the region near the first HXR peak of the flare in EUV images taken around 07:30:10~UT.
Figure \ref{totalarea} shows the DEM divided by the number of pixels in the area as a function of temperature from each of the four
selected regions. The DEM from all of the regions suggests the presence of two temperature components,
a low-temperature component around 2 MK, and a high temperature component around/above 8 MK. There are several possible
reasons for this bi-model distribution: (a) line of sight effect. The low temperature component may not originate from the flaring loop
itself, but from higher layers of the atmosphere; 
(b) true multi-temperature structure of the flaring site; (c) artifacts related to the SDO/AIA response calibration or the assumed theoretical spectral calculations. 
To test for (a) we calculated the DEM for the same four regions in images taken around 15 minutes, 10 minutes, 5 minutes, 2 minutes and 1 minute before the analyzed flare time interval (Figure~\ref{preflaredem}).
The time evolution for all four flare regions suggests that the low temperature component is always present,
while the high temperature component increases in intensity by more than one order of magnitude during the rise phase of the flare.

The second possibility (b) can be tested by analyzing the pixel by pixel DEM parallel as well as perpendicular to the magnetic loop. Such analysis suggests that the intensity of the low temperature component is constant both in parallel and perpendicular direction to the magnetic field, while the intensity of the high temperature component in the parallel direction in e.g. the coronal source region decreases by a factor of 3 at larger heights consistent with reduced flare emission above the top of the coronal source. This is supported by images in 193 \AA\ that show faint loop structures which are probably not associated with the flare. Thus, it is unlikely that the low temperature component is due to flaring emission but it can be explained as line of sight effect. In other words, the high temperature component dominates the actual flare emission and this component is what we focus on.

From the DEM curve in Figure~\ref{totalarea} we can then calculate the total emission measure per area.
Here we define the total emission measure as the DEM integrated over the main (hot) flare temperature component from $T_{min}=2.5-3$~MK to $T_{max}=32$~MK (see Figure \ref{totalarea}):
\begin{equation}
\frac{EM}{A}=\int_{T_{min}}^{T_{max}} \xi(T) \mathrm{dT}
\end{equation}
where $A$ is the area of the analyzed region $A=(0.6\times 7.25\times 10^7)^2\times N_{pix}$ $\mathrm{cm^2}$ with $N_{pix}$ being the pixel number over which the flux was averaged.
For comparison with RHESSI and GOES we define a peak, or main, temperature as the first moment of the DEM curve between $T_{min}$ and $T_{max}$:
\begin{equation} \label{teq}
<T>=\frac{\int_{T_{min}}^{T_{max}}\xi(T)\times T\mathrm{dT}}{\int_{T_{min}}^{T_{max}}\xi(T)\mathrm{dT}},
\end{equation}
and the second moment gives an uncertainty range of possible temperatures. All temperatures are given in the form $<T>=<T>\pm \Delta T$, where $<T>$ is the temperature found from Equation~\ref{teq} and $\Delta T$ is the half width of the second moment. 
In the next sections we discuss the results for each of the four regions individually. Table \ref{table} summarizes the temperature and emission measure values.
\subsection{Soft X-ray coronal source}
The SXR emission originated from higher altitude than the EUV loop as seen in 94~\AA, but coincided with the 193~\AA\ emission.
The SXR thermal parameters (temperature and emission measure) of the coronal source were found by fitting the full Sun RHESSI spectrum which is dominated by the coronal source emission at low energies. A thermal component plus a single power-law thick target component with a low energy cutoff were fitted over a time-interval of 07:30:00 - 07:30:40 UT (same time interval as RHESSI imaging). Fits for two different values of the low energy cutoff resulted in comparable goodness of the fit, thus we cite the two resulting values of the temperature and emission measure, which also gives a confidence interval for these parameters. The temperature resulted in $T_{RHESSI}=18.5\pm 1.5$~MK and the emission measure $EM_{RHESSI}=(9.4\pm 3.4)\times 10^{47}\,\mathrm{cm^{-3}}$, respectively. For comparison with AIA emission measures, the emission measure per unit area $EM_{RHESSI_A}$ is used. Here we assume that the RHESSI emission originated from an area the size of the 50\% contours in a CLEAN image ($8.5\times 10^{17}$~cm$^2$, CLEAN beam corrected), therefore $EM_{RHESSI_A}=(1.1\pm 0.4)\times 10^{30}\,\mathrm{cm^{-5}}$.  Another method is to use filter ratios from the GOES satellites \citep{White95} where we assume the same area as for RHESSI. This results in an emission measure per area of $EM_{GOES_A}\sim 3.1\cdot 10^{30}\,\mathrm{cm^{-5}}$ at a temperature of 14~MK. The EUV emission measure and temperatures were found as described above and amount to $6.8\pm 1.3$~MK and $EM_A=(5.8 \pm 1.1)\times 10^{28}$~$\mathrm{cm^{-5}}$.
\subsubsection{Saturation and high temperature wavelength channels} \label{secsaturation}
\begin{figure}
\begin{center}
\resizebox{0.95\hsize}{!}{\includegraphics{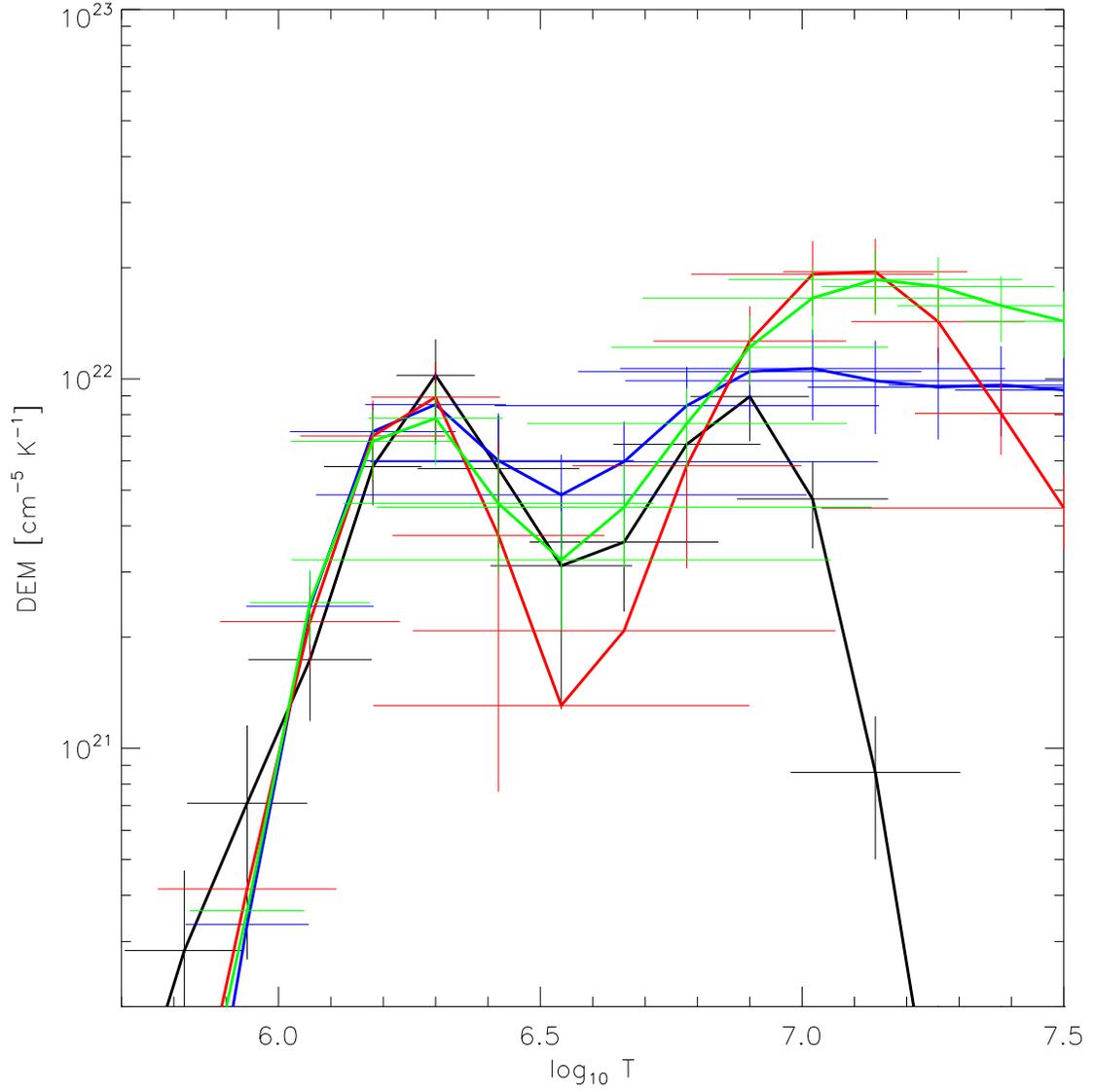}}
\end{center}
\caption {DEM of region 1 (black). The blue and red lines give the DEM when the 193 \AA\ and 131 \AA\  wavelengths are included in the analysis despite saturation. Green is the DEM when both, 193 \AA\ and 131, \AA\ are included. }
\label{193}
\end{figure}

The 193~\AA\ and 131~\AA\ images were saturated near the SXR coronal source and across the whole loop, respectively, which is why they cannot be used to find the DEM in some regions. However, the images suggest that the 193~\AA\ emission was co-spatial with the RHESSI coronal source. Since the 193~\AA\ response has a peak at around 16 MK \citep{Le11}, which is close to the observed RHESSI temperature, it is likely that the two images show the same emission at similar temperature. Thus, by omitting the 193~\AA\ and 131~\AA\ wavelength channels in the DEM calculation, most high temperature emission will be missed. To estimate the effect on the DEM we calculated DEMs by including 193~\AA\ and 131~\AA\ and compared the resulting DEM and total emission measure per area. Figure \ref{193} shows this comparison in the case of exclusion of 193~\AA, 131~\AA\ individually and both at a time. When these wavelength channels are included the emphasis is on the high temperature emission, the temperature range extends to the upper limit in the DEM reconstruction (32 MK) and the emission measure per area is found as $EM_A=(4.2 \pm 1.1)\times 10^{29}$~$\mathrm{cm^{-5}}$ which is 7 times higher than when these wavelength channels are not used but is still 2.6 times smaller than the RHESSI emission measure (compare Table \ref{table}). However, the data values from the 193~\AA\ and 131~\AA\ images are lower limits due to the saturation. This points out the effect of omitting the high temperature response wavelength channels and the obtained coronal source emission measure has to be viewed as a lower limit. Furthermore the temperature sensitivity range of AIA has to be considered as additional limiting factor. Temperatures around 20 MK are at the upper limit of the AIA sensitivity (the high temperature response of 193~\AA\ peaks at 16~MK). Therefore, if the bulk of the plasma is at such high temperatures it is likely that the DEM analysis will result in a smaller emission measure due to lack of sensitivity, even if all coronal wavelength channels are used. 
\subsection{EUV loop}
The EUV loop was divided into two regions, one just below the SXR coronal source (region 2) and one closer to the HXR footpoint (region 3).  The temperature and emission measure in region 2 were $7.5\pm 2.7$~MK and $(1.2\pm 0.3)\times 10^{29}$ $\mathrm{cm^{-5}}$, respectively. In region 3, values of $7.4\pm 2.2$~MK and $(2.1\pm 0.6)\times 10^{29}$~$\mathrm{cm^{-5}}$ were found. Thus the temperature is constant within the uncertainty along the loop while the emission measure seems to decrease slightly with increasing height. We note that due to line-of-sight effect the emission could partially originate from the loops behind/in front of the loop under study. 
\subsection{Footpoint}
Near the position of the HXR footpoint (region 4) the temperature was $8.7\pm 2.0$~MK, while the emission measure per area was found as $(2.1\pm 0.8)\times 10^{29}$~$\mathrm{cm^{-5}}$. Note that this region, though partly overlapping with the HXR footpoint emission in images, corresponds to a height of $\approx$ 2.2 - 2.9 Mm. This is above the height of the peak HXR emission ($\sim$ 1.6 Mm at 30 - 40 keV following \citet{Ba11b}) and above the height of the WL emission which is co-spatial with the HXR emission \citep{2012ApJ...753L..26M}. Thus, the emission analyzed here originates from the upper transition region / lower corona rather than the chromosphere (compare Figure \ref{saturation}).
\subsection{Comparison of thermal parameters and density}
In this section we compare the thermal parameters found from EUV data with the RHESSI observations, as well as with GOES measurements, and we also determine the densities.
In a next step, every radial ($x$)-coordinate is mapped to a height above the photosphere, using the reference coordinate that represents the photospheric height found from visibility analysis of RHESSI footpoints as described in \cite{Ba11a,Ba11b}, where we found a reference position of $r_0=929.4\pm0.3$~arcsec.

Similar to previous results \citep[e.g.][]{Ba05,Ha08} the RHESSI temperature is highest, while the GOES emission measure is higher but at a temperature a few MK lower than the RHESSI temperature. The EUV emission measures per area from all four regions of interest are up to two orders of magnitude smaller than the RHESSI and GOES emission measure. This can be explained by a number of reasons. The RHESSI and GOES emission measures are calculated for the full Sun, though in the case of a flare this emission originates predominantly from the flaring site. To compare the emission measure per area with the value from AIA, we assume that the bulk of the RHESSI and GOES emission originates from within an area corresponding to the 50\% contours in a RHESSI 6-12 keV image. We can calculate the AIA emission measure of the coronal source for the same area to find $EM=(1.2\pm 0.3)\times 10^{29}$ $\mathrm{cm^{-5}}$. This is a factor of two larger than the emission measure from region 1, but still almost one order of magnitude smaller than the RHESSI emission measure.
The main reason, as illustrated in Section \ref{secsaturation}, is that emission at highest temperatures is missed by omitting the highest temperature response wavelengths 193 and 131~\AA. This is likely worsened by the fact that the temperature at this stage of the flare is higher than the range to which AIA is most sensitive, even if the high temperature response was included, thus the emission measure is smaller. \\

\begin{figure}
\begin{center}
\includegraphics[height=5cm]{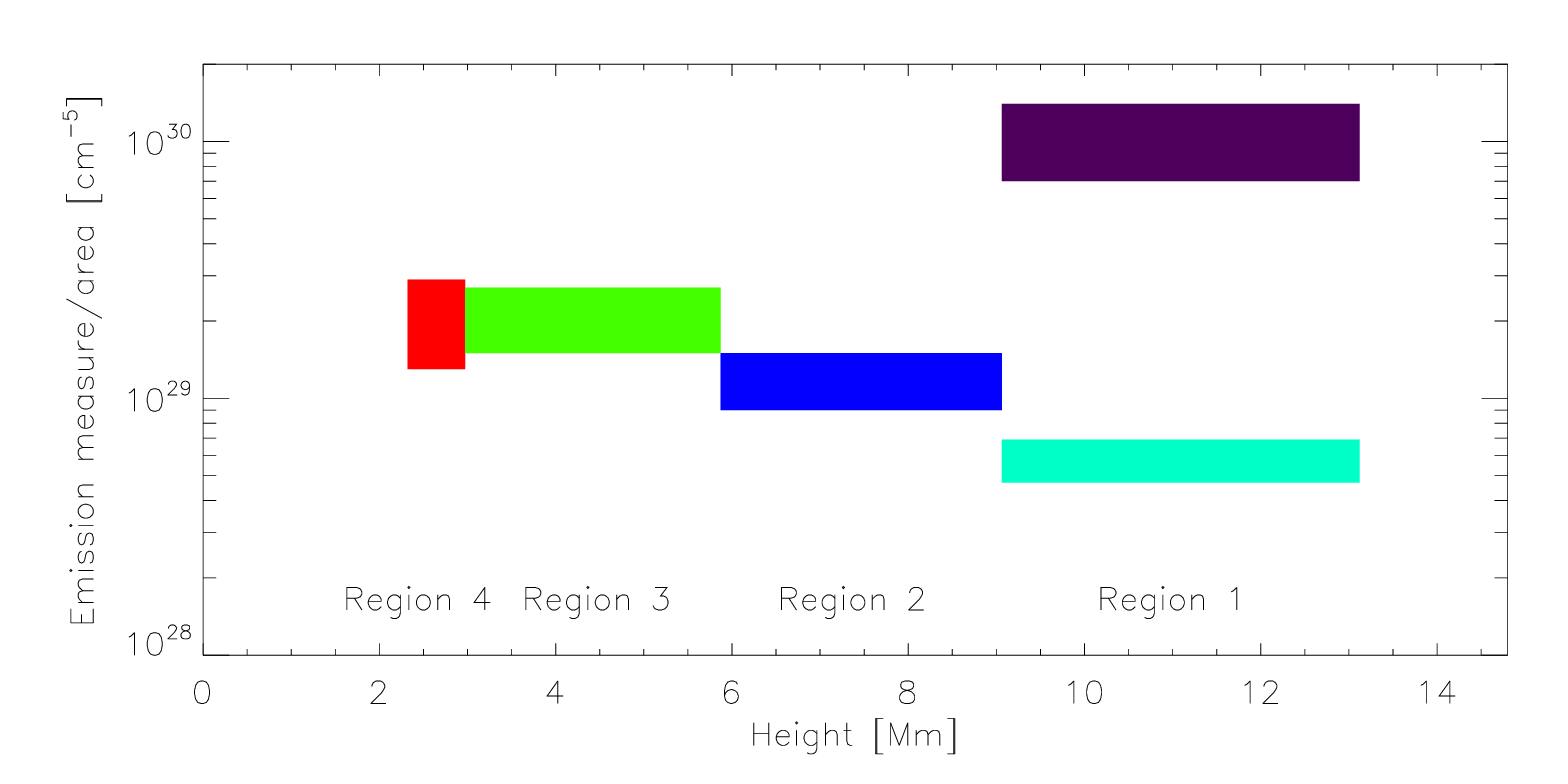} \\
\includegraphics[height=5cm]{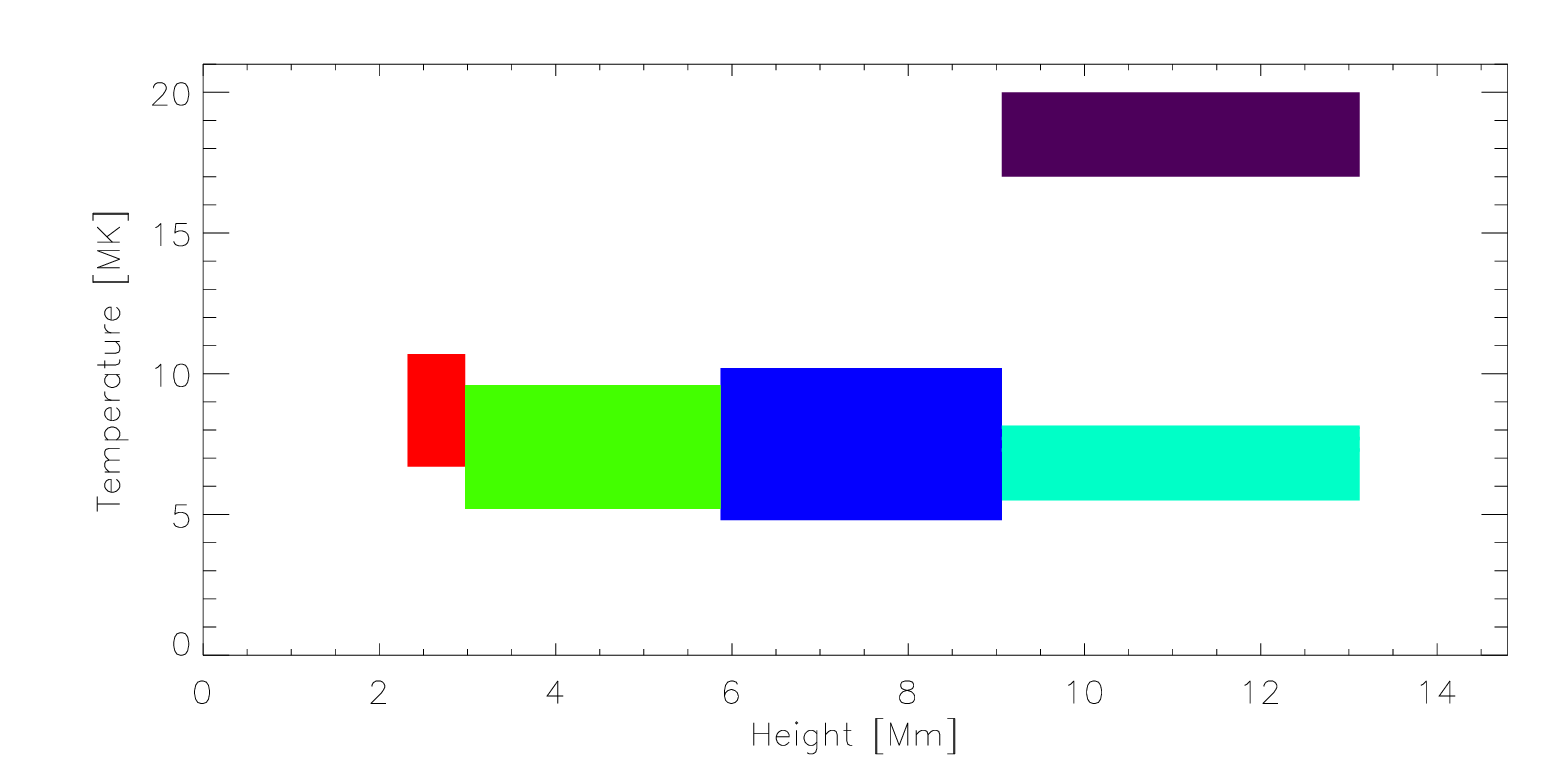} \\
\includegraphics[height=5cm]{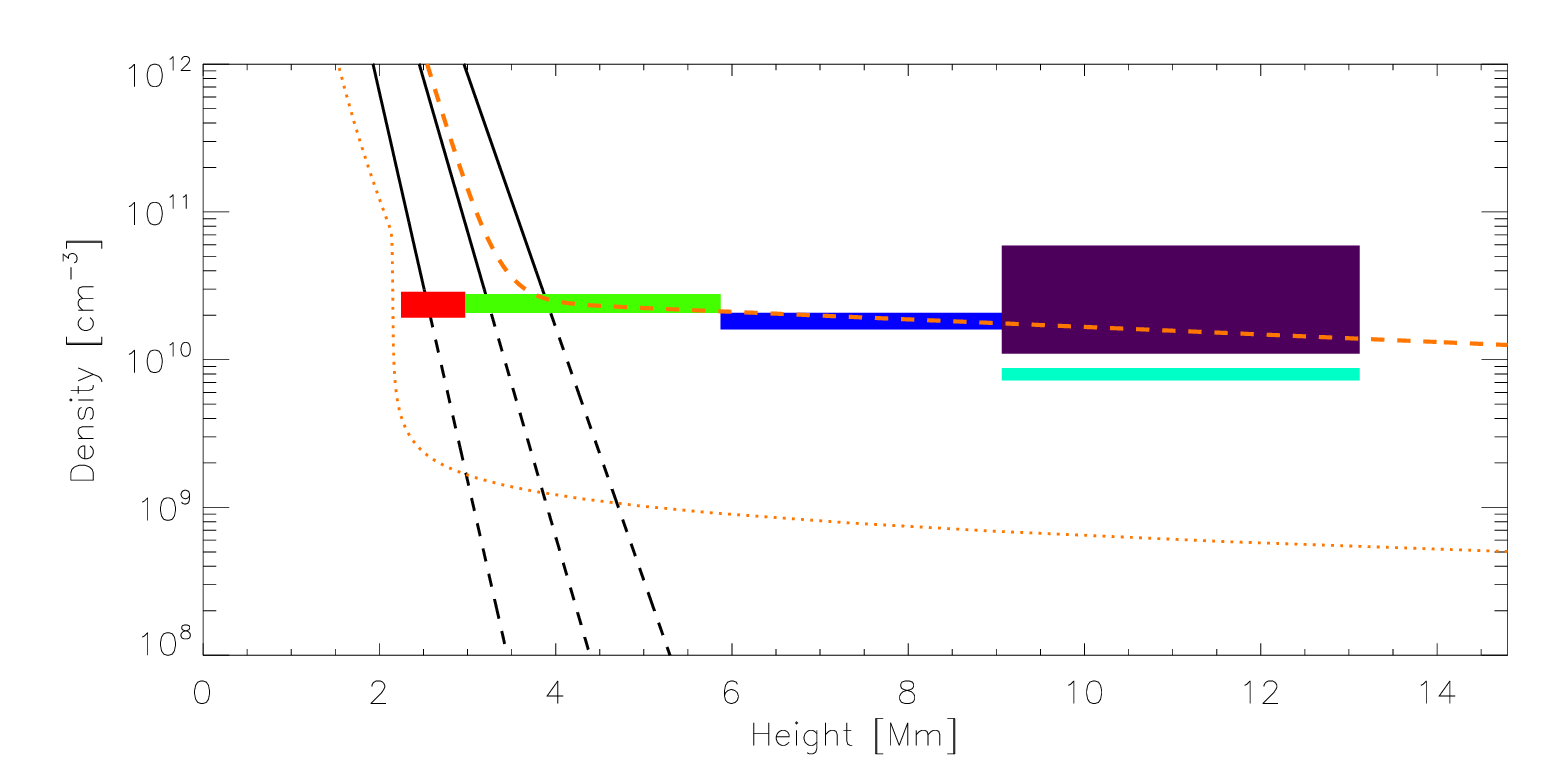}
\end{center}
\caption {Top: Emission measure per area as a function of height found for the different regions and using different methods. Red: Region 4, green and blue: Regions 3 and 2 turquoise: coronal source from AIA (region 1), purple: coronal source from RHESSI. Middle: Temperature. Bottom: Densities. The black lines give the footpoint density model and its uncertainties as found via HXR visibility analysis (taken from \citet{Ba11b}). The dashed line gives the sum of two exponential density functions with scale-heights of 220 km and 17000 km, respectively. The dotted line indicates the quiet Sun model by \citet{2008ApJS..175..229A}.}
\label{densities}
\end{figure}

In a next step we compare the densities found from the different methods.
An estimate for the density of the coronal source in SXR can be found by measuring the area $A_{cs}$ of the coronal source in CLEAN or Pixon images and using $n_{cs}=\sqrt{EM_{cs}/V_{cs}}$, where the volume is estimated to be $V_{cs}=A_{cs}^{3/2}$. In this case, using 50\% contours in a CLEAN image at 6-12 keV and accounting for the CLEAN beam, we get a volume of $7.9\times 10^{26}\,\mathrm{cm^3}$ and thus a density of $n_{cs}=(3.4 \pm 0.6) \times 10^{10}\,\mathrm{cm^{-3}}$. In the case of AIA observations the density is given as the square root of the emission measure per area divided by the distance along the line of sight. For the line of sight component the observed portion of the loop is assumed to be `''cylindrical''. We use the size of the HXR footpoint to estimate the loop diameter and approximate the line of sight distance to 5~arcsec = 3.6~Mm for regions 2-4. The coronal source (region 4) appears extended compared with the loop, therefore a value of 9~Mm is used for the line of sight distance. It is often not possible to determine the thermal parameters of the footpoints, even in imaging spectroscopy \citep{Ba06,1994ApJ...422L..25H,1993ApJ...416L..91M}, either because of the dominance of the coronal source and the dynamic range of RHESSI, or because the plasma is at temperatures to which RHESSI is not sensitive. However, using RHESSI visibility analysis it is possible to determine the chromospheric density near the footpoints by fitting an exponential density model to the energy dependent positions of the footpoints \citep{Ba11b,Koet10,Ko08}. Figure \ref{densities} gives an overview of the densities as a function of height. The densities in regions 2-4 are all around $2\times 10^{10}$~$\mathrm{cm^{-3}}$. A similar density is found using the SXR RHESSI data. This is consistent with previous densities found from RHESSI thermal analysis and indicates a constant density along the loop. For these calculations a filling factor of one was assumed. It cannot be excluded that the filling factor is smaller. However, recent analysis of several flare loops suggest a value that is generally close to one \citep{2012ApJ...755...32G, 2012A&A...543A..53G}. Even a filling factor as low as 0.1 would lead to a density only 3 times higher which is still feasible for a flaring loop. 

Since the density found from HXR observations is the neutral plasma density, and AIA observes hot, ionized plasma, combining these observations gives a complete picture of the density in the solar atmosphere during a flare. The combined observations represent a steeply decreasing density from the photosphere to a height of about 4-5 arcsec with a flatter decrease or near constant value at larger heights, similar to models such as from \citet{Ver81,2008ApJS..175..229A}. The \citet{2008ApJS..175..229A} quiet Sun model is shown in Figure \ref{densities} along with a double-exponential density model with two scale-heights (220 km at small heights, 17000 km at larger heights) that follows the observed pattern. The AIA-determined density of region 1 is lower by about a factor of 4 compared with RHESSI because of the small emission measure. 
\section{Discussion and Conclusion}
In this work we present a comprehensive study of the thermal parameters (emission measure, temperature) and the density in four different regions of a flaring loop, using both X-ray as well as EUV data.
Detailed temperature, emission measure and density analysis along the loop near the flare HXR peak indicates that the hottest plasma is found near the coronal loop top source, in line with previous observations. The EUV observations suggest that the density in the loop legs increases with decreasing height, but the temperature is constant within uncertainties. The footpoint structure is such that the plasma is neutral or partly ionized up to $\sim$ 1-2~Mm with a steep, unresolved rise to $\sim$ 5~MK 1-2~Mm above the height of the HXR and WL emission. 

The DEM in EUV was found using the method of regularized inversion developed by \citet{2012A&A...539A.146H}.
We show that the method is capable of recovering two temperature components, one at around 2~MK and one at or around 8 MK. The low temperature component around $2$~MK is observed in all 4 regions and is only weakly changing with time, while the intensity of the high temperature component increases by more than one order of magnitude at the time of the flare. Therefore, the low temperature component can be attributed to line of sight effect. The existence of this component has to be kept in mind when DEMs are determined by fitting single temperature Gaussians. 
The results also show the presence of hot ($\sim > 8$~MK) plasma not only in the coronal source but also in the coronal loop legs, and near the footpoints of the loop.
As the flare progresses, the high temperature emission starts to increase from around $07:15$~UT with the emission finally saturating the `hottest' wavelength channels around $07:30$~UT near the first HXR peak. This rise is first seen in the coronal source and then propagates downward to the footpoints. This is consistent with the start of the flare and initial heating taking place in the corona.
Note also that the loop top emission in the SDO/AIA image in 193~\AA\ is co-spatial with RHESSI coronal loops, while the other ``hot'' wavelength channel 94~\AA\ is not. The loops visible in 94~\AA\ are around $5''$ lower than the 193~\AA\ loop structure. While the brightest soft X-ray feature seen by RHESSI in $10-15$~keV is undoubtedly the coronal source, the brightest 193~\AA\ source is in fact the region to the north of the flare located near $x=-925''$,$y=285''$. This is intriguing since the emission seems to be part of the active region and as such contains a significant fraction of the thermal energy. However, this region is not associated with either RHESSI source at any stage of the flare. 

The saturation in the ``hottest'' wavelength channels 193~\AA\ and 131~\AA\ disallows the study of the hottest plasma and limits the temperature range of which plasma can be analyzed. 
Comparison of the total AIA emission measure with RHESSI and GOES emission measures suggests that the EUV emission is up to more than one order of magnitude smaller than the X-ray emission measure. This can be partly attributed to omitting the highest temperature wavelength channels in the analysis due to saturation, and it emphasizes the importance of using all 6 ``coronal'' wavelength channels for the DEM reconstruction. An additional likely cause it that the bulk of the plasma is at a temperature higher than the main sensitivity range of AIA. Cooler events (i.e. with a RHESSI temperature closer to 10~MK) are probably better suited for more detailed comparative studies. In addition, in weaker events the likelihood for AIA saturation is reduced. 

Combined RHESSI/SDO-AIA studies allow to investigate in detail a broad temperature range from 1-2~MK up to $\sim$ 20~MK. Moreover, the use of DEM-analysis helps to quantify the temperature, emission and density structure in the flaring loop and fills the gap between the coronal source and footpoints that are observed in SXR and HXR.

\acknowledgments
Financial support by the STFC UK rolling grant, the Leverhulme Trust,
and by the European Commission through HESPE (FP7-SPACE-2010-263086)
and the "Radiosun" (PEOPLE-2011-IRSES-295272) Networks
is gratefully acknowledged. MB was supported by a travel grant from the Swiss Society for 
Astronomy and Astrophysics (SSAA). We would like to thank Iain Hannah and S\"am Krucker for helpful discussions and James Lemen for useful comments.

\bibliographystyle{apj}
\bibliography{mybib}


\begin{deluxetable}{lllc}
\tabletypesize{\scriptsize}
\tablecaption{Temperatures, emission measures, and densities found by the different methods}
\tablehead{
\colhead{Instrument/region} & \colhead{T [MK]} & \colhead{EM/area [$\mathrm{cm^{-5}}$]} & \colhead{$n_e$ $\mathrm{[cm^{-3}]}$} }\startdata
RHESSI/ coronal source & $18.5\pm1.5$ & $(1.1\pm 0.4)\times 10^{30}$ &  $(3.4\pm 0.6) \times 10^{10}$\\
AIA/ Region 1 &$6.8\pm 1.3$  &  $(5.8\pm 1.1)\times 10^{28}$&$(8.0 \pm 0.8)\times 10^{9}$\\
AIA/ Region 1 (incl. 193 \AA\ ) &$11.6\pm 3.9$  &  $(2.7\pm 0.9)\times 10^{29}$&$(1.7 \pm 1.0)\times 10^{10}$\\
AIA/ Region 1 (incl. 131 \AA\ ) &$12.4\pm 3.0$  &  $(3.2\pm 0.8)\times 10^{29}$&$(1.9 \pm 0.9)\times 10^{10}$\\
AIA/ Region 1 (incl. 131 \& 193 \AA\ ) &$13.6\pm 4.0$  &  $(4.2\pm 1.1)\times 10^{29}$&$(2.2 \pm 1.0)\times 10^{10}$\\
AIA/ Region 2 &$7.5\pm 2.7$&$(1.2\pm 0.3)\times 10^{29}$&$(1.8\pm 0.2)\times 10^{10}$\\
AIA/ Region 3 &$7.4\pm 2.2$&$(2.1\pm 0.6)\times 10^{29}$&$(1.5\pm 0.2)\times 10^{10}$\\
AIA/ Region 4 &$8.7\pm 2.0$&$(2.1\pm 0.8)\times 10^{29}$& $(2.4\pm 0.5)\times 10^{10}$\\
GOES/coronal source & 14 & $3.1\times 10^{30}$& \\
\enddata
\label{table}
\end{deluxetable}


\end{document}